\documentclass[notitlepage, 11pt]{article}

\usepackage[myheadings]{fullpage}
\usepackage{amsmath}
\usepackage[table,xcdraw]{xcolor}
\usepackage{graphicx}
\usepackage{float}
\usepackage{threeparttable}
\usepackage{setspace}
\usepackage[toc,page]{appendix}
\usepackage{csquotes}
\usepackage[natbibapa]{apacite}
\usepackage{natbib}
\renewcommand{\vec}[1]{\mathbf{#1}}

\usepackage{algorithm}  
\usepackage{algorithmic}  

\renewcommand{\vec}[1]{\mathbf{#1}}

\title{Modeling Heterogeneous Peer Assortment Effects using Finite Mixture Exponential Random Graph Models}
\author{Teague R. Henry, Kathleen M. Gates, Mitchell J. Prinstein \\ \textit{University of North Carolina at Chapel Hill}\\ Douglas Steinley \\ \textit{University of Missouri}}

\date{9/4/2019}

\begin{document}

\begin{titlepage}

\maketitle
\begin{abstract}
This article develops a class of models called Sender/Receiver Finite Mixture Exponential Random Graph Models (SRFM-ERGMs). This class of models extends the existing Exponential Random Graph Modeling framework to allow analysts to model unobserved heterogeneity in the effects of nodal covariates and network features without a block structure. An empirical example regarding substance use among adolescents is presented. Simulations across a variety of conditions are used to evaluate the performance of this technique. We conclude that unobserved heterogeneity in effects of nodal covariates can be a major cause of mis-fit in network models, and the SRFM-ERGM approach can alleviate this misfit. Implications for the analysis of social networks in psychological science are discussed.

\end{abstract}



\end{titlepage}

\setcounter{page}{2}


\section{Introduction}
Psychologists and social scientists increasingly utilize methods to analyze social networks. These methods attempt to quantify how individuals' behaviors or traits relate to specific peer selection processes, thus depicting how a given individual’s behavior or trait changes the network structure over time. The cross-sectional pattern of peer nominations conditioned upon individual traits and covariates is called ``peer assortment". The relationships between covariates and peer assortment present complex methodological problems.  One such problem is that heterogeneity in the peer selection and influence processes likely exists. Specifically, some individuals may have different patterns of relations among behaviors, traits, and peer assortment. The present paper introduces a novel method that utilizes finite mixture modeling within an exponential random graph modeling framework to allow for heterogeneity in the patterns among covariates and peer assortment.  

In recent years, new methodologies have been developed to analyze peer selection processes (e.g., SIENA; \citealp{snijdersintroduction2010}; TERGMs; \citealp{Hanneke2010}; STERGMs; \citealp{Krivitsky2014}).  Models that investigate cross-sectional peer assortment are similar to models that assess peer selection processes, specifically the exponential random graph modeling framework  \citep[ERGM;][]{frankmarkov1986,wassermanlogit1996}. What underlies all of these network analytic methods is their ability to assess the effect of individual level traits and covariates, such as personality characteristics or observed behaviors, on the network structure.  In directed networks (e.g., peer friendship networks where individuals nominate peers as friends, but there is no requirement for bidirectional nomination), researchers can assess the effects of covariates on nominating a peer (sending) or being nominated (receiving). These effects assess the impact of a behavior, such as substance use, on the tendency for individuals to send or receive edges (i.e., nominate or be nominated). Identifying covariate effects of sender and receiver components prove useful when investigating the impact of individuals’ behaviors on network structure. 

Parallel to network methods’ recent flourishing, methods that seek to model and capture individual (or ideographic) differences are also resurging and may benefit from a network approach. Modeling ideographic differences can be done a variety of different ways. For example, an ideographic analysis of relations among variables \citep{molenaara2004} allows researchers to examine a single individual's pattern of behavior, while other methods test the aggregate relations between outcomes as moderated by some individual trait or behavior \citep{bauerevaluating2011}. Of the existing individual differences methods, moderation by an observed variable is by far the most commonly used \citep{bauerevaluating2011}. Indeed, neglecting to include moderation effects in non-network data may result in biased estimates \citep{jaccardinteraction2003} and  spurious random effects \citep{bauerconsequences2009}. It follows that failure to account for heterogeneity in network analysis could have potentially severe consequences, especially because certain parameter combinations can cause the estimation to converge on a degenerate solution \citep{chatterjeeestimating2013}. This risk of degeneracy could potentially increase when estimating a mis-specified model.

Given burgeoning interests in networks and individual differences, a growing number of researchers are seeking to examine the role of moderators and individual differences in networks (for review see, \citealp{brechwaldbeyond2011}), with the majority of attention going to peer influence and little work exploring potential moderators of peer assortment (with exception; see \cite{Kiuru2010}). Despite the interest in the field, there are a limited number of methodological approaches within network science for moderation of these effects based on unobserved classes of nodes. There are several statistical methods for detecting latent communities, such as stochastic block modeling \citep{Nowicki2001}, latent space modeling \citep{hofflatent2002}, a latent space model that allows for covariate effects \citep{Handcock2007}, a mixture latent community approach \citep{Daudin2008}, and more recently, the ERGM approach \citep{schweinbergerlocal2015} that allows for differences in the effects of covariates within a community, as well as a stochastic block modeling approach that also allows for covariate effects separate from a block structure \citep{Sweet2015}. However, all of these approaches focus on assessing latent community structure and do not focus on assessing individual differences in the effects of covariates based on unobserved subgroups. Clusterwise $p^*$ modeling \citep{Steinley2010} also focuses on determining latent community structure, but has the feature of using an ERGM model to inform the assignment of nodes to the latent communities. The method developed in this manuscript bears resemblance to the clusterwise $p^*$ model, however does not require that the latent class structure result in interconnected latent communities. In this way it can be looked at as a generalization of the work in \citet{Steinley2010}.

\subsection{Community Structures Vs. Latent Classes}

One key difference between previously developed models and the latent class framework presented here is the focus on detecting different \textit{classes} of nodes/actors within the network, where the classes do not correspond to community structures (i.e. sets of nodes that are highly interconnected). For example, the ERGM approach of \citet{schweinbergerlocal2015} requires the existence of cohesive communities of nodes. Within each community, a separate ERGM model is estimated, while between community, dyadic independence ERGM models are fit. \citet{schweinbergerlocal2015}'s modeling framework allows for community dependent covariate effects, however this in turn implies that all members of a particular community have the same covariate effect. The work of \citet{Tallberg2004} presents a stochastic block model that allows for nodal covariates, as well as have those nodal covariates predict community membership, but this model does not allow for community specific covariate effects. Relatedly, the covariate stochastic block model of \citet{Sweet2015} allows for the estimation of covariate effects independent of a co-occurring block membership, but does not allow for community specific covariate effects.

These community based covariate models are a powerful set of techniques for identifying cohesive communities of nodes and estimating the specific effect of covariates within each community. However, as these methods were primarily driven by the desire to detect community structures, they are less able to identify subsets of nodes that differ on their covariate effect, but do not take part in any sort of block structure. Take for example the effect of substance use on friendship formation for adolescents. It is not inconceivable that there would be sets of adolescents who use substances in a non-social setting, and therefore would have no relation between the amount of substance use and number of friends, while in a different set of adolescents, substance use might be social, and for these individuals, substance use might correlate highly with number of friends. Importantly, these differential effects do not necessarily lead to a community structure. To differentiate between community-type models and models that do not need a community, throughout this manuscript we use the term \textit{latent class} to denote node membership in an unobserved grouping that does \textit{not} imply a community structure. The choice between estimating a community structure and estimating latent classes in network data has important consequences for estimation, and the differences between the two types of models with regards to estimation will be discussed further below. It is important here to note that community structure models are indeed examples of latent class models more generally, as both are seeking to partition nodes into a set of equivalence classes. There have been examples of these latent class ERGMs in prior literature. Specifically, the latent variable ERGM of \citet{Koskinen2009} presents a Bayesian approach to estimating heterogeneity in covariate effects. However, this approach is not scalable to many classes or large networks, due to the MCMC estimator employed. 

One additional type of model with relevance to the unobserved heterogeneity in covariate effects is that of ERGMs with nodal random effects \citep[][e.g.]{Zijlstra2006,Duijn2004a,Thiemichen2016}. This class of model assumes that nodal sender and receiver effects (i.e. a node's overall tendency to send/receive ties) is normally distributed rather than a single fixed estimate. As originally proposed, these models do not contain random effects for covariates, but in theory this would be a simple extension. While these types of models are ideal for allowing heterogeneity between nodes with respect to their sender/receiver intercepts, these models do not allow for starkly different classes of nodes, and are therefore not appropriate for the detection of sub-types within a social network. 

The present manuscript aims to extend the definition of mixtures on networks, and model mixtures of individuals who differ in their effect of observed variables on their position in the network. Latent class models and mixture models, which are also commonly used in psychological research to assess individual differences based on unobserved data, may provide a solution to this methodological hindrance. Specifically, finite mixture linear regression \citep{desarboa1988} and finite mixture general linear regression \citep{wedela1995} can be used to partition a sample into groups, in which each group has different relations among observed variables.  Following this line of research, this manuscript develops a sender/receiver finite mixture modeling approach for ERGMs (SRFM-ERGM), focusing on cross-sectional peer assortment models. This work will help address the methodological conundrum facing researchers who wish to utilize a whole network analysis method to test the moderation of peer assortment by individual differences. We first present sufficient details of ERGMs as well as the methodological innovation. Next, we present a motivating empirical example examining individual differences in the effect of alcohol use on the network structure in a middle-school will be presented.  Finally, we present a set of simulations that demonstrate the SRFM-ERGMs ability to recover class membership, as well as the consequences of failing to model the individual differences.

\section{Network Terminology and Model Description} 

As social networks typically take the form of binary (in that a social relationship is either present or not), and directed (where a social relationship is not necessarily reciprocated), the description of network terminology will be specific to binary directed networks. A network, $g$, consists of a set of {\it nodes} and a set of \emph{edges} that connect the nodes. Nodes define the members of a network and, in the case of social networks, usually represent people. Edges define the relationship between nodes in the network. In the case of social networks, this is a social relation, such as friendship. In a directed network, relations do not need to be reciprocated, as individual $i$ might nominate individual $j$, while individual $j$ might not nominate individual $i$. A common means of representing a network is that of an adjacency matrix. An adjacency matrix for a binary, directed network $g$, \textbf{A}, is a $N \times N$ matrix where where $N$ is the number of individuals in that comprise the network and $a_{ij}$ is $1$ if individual $i$ has nominated individual $j$ and $0$ otherwise. 

In addition to the presence or absence of edges, often times \emph{nodal covariates} (such as gender or substance use in social networks) are also included in the analysis.  In this manuscript, the $N \times P$ matrix of node level variables will be denoted using $\mathbf{X}$.

\section{Exponential Random Graph Models}

ERGMs, also referred to as $p^*$ models \citep{wassermanlogit1996}, are a family of models used to analyze network structure both contingent on structural predictors such as number of triangles or reciprocity of edges as well as on covariates of the individual nodes. ERGMs model the entire network as a sample from a population of networks with a set of sufficient statistics associated with the distribution. These sufficient statistics can include the number of edges in the network, the number of edges between individuals with dissimilar values on one specific covariate and/or the number of closed triangles in the network, to name a few examples. ERGMs were originally developed to analyze {\it dyad independent} networks, where the probability of an edge between two nodes was only dependent on the characteristics of those two nodes and reciprocity in edges between the two given nodes \citep{Holland1981b}. They were extended by \citet{frankmarkov1986} to include transitivity effects, where the probability of an edge existing depends on the existence of neighboring edges. The modeling of transitivity proves to be difficult, and only in recent years have successful modeling techniques been developed \citep{Hunter2006}. 
        
An ERGM models the observed network as a sample from a exponential family distribution with the following form.
Let $G$ be a random directed network of size $N$ with sample space $\mathcal{G}$, and $\mathbf{X}$ a $N \times P$ matrix of fixed and known nodal covariates. $g$ is then the sample realization of $G$, which has a corresponding adjacency matrix $\mathbf{A}$, with $a_{ij}$ referring to the value of the edge between nodes $i$ and $j$, and equivalently the $ij$th element of the matrix $\mathbf{A}$. Then the probability density function of $G$ is
\begin{equation}\label{eq:ergm}
P(G = g|\mathbf{X}, \boldsymbol{\theta}) = \frac{\exp[\langle\boldsymbol{\theta},s(\mathbf{a}, \mathbf{X})\rangle]}{\psi(\boldsymbol{\theta})}
\end{equation}
where $\boldsymbol{\theta}$ is a $d$ length vector of natural parameters in log-odds metric, $s(\mathbf{A}, \mathbf{X})$ is a $d$ length vector of sufficient statistics, $\langle\boldsymbol{\theta}, s(\mathbf{A}, \mathbf{X})\rangle$ is the inner product of the vector of natural parameters and the vector of sufficient statistics and $\psi(\boldsymbol{\theta})$ is a normalizing constant. The log-likelihood of the parameters \cite{Hunter2006} is 
\begin{equation}\label{eq:llergm}
\ell(\vec{\theta}) = \langle\boldsymbol{\theta},s(\mathbf{A}, \mathbf{X})\rangle - \log(\psi(\boldsymbol{\theta})).
\end{equation}

As was outlined above, the vector of sufficient statistics $s(\mathbf{a}, \mathbf{X})$ can contain a variety of terms.  Broadly, these can be divided into two categories, sufficient statistics that are wholly based on the network structure, {\it structural statistics}, and statistics that are based on the nodal covariates. Sufficient statistics that will be used in the simulation studies and the empirical examples are described later in the manuscript.

 \section{Estimation for ERGMs}
 
Estimation of the parameter vector $\boldsymbol{\theta}$ for an ERGM model is complicated when a term that measures transitivity is included in the model.  ERGMs, as originally proposed by \citet{Holland1981b}, did not contain transitivity terms, and instead were considered  dyad independence models. Dyad independence models are ones where the state of a dyad (what edges in a dyad are present) is conditionally independent of all other dyads given the characteristics of the nodes that make up the dyad in question. This simplifies the likelihood of a {\it dyad independent} ERGM to that of a multinomial regression for a directed network (the categories being the 4 possible states of a directed dyad), or a binary logistic regression for an undirected network, which in turn makes estimation of a dyad independence ERGM trivial.

When a term that accounts for transitivity, such the count of closed triangles, is added to the model, the resulting dependency structure complicates estimation.  Specifically, the likelihood changes due to a normalizing constant that is defined with a sum over all possible networks of a given size, which is an astronomically large number for any reasonable sized network. To illustrate, the estimated number of stars in the observable universe is roughly between $10^22$ and $10^23$. A network of directed friendship nominations of $100$ individuals has $2^100$ possible configurations. The number of possible configurations of that network is approximately $12676506$ times larger than the estimated number of stars in the observable universe, which makes calculating the normalizing constant somewhat difficult. 

A great deal of work has been done on developing approximate MLE estimators for ERGM models, and one commonly used estimator is the MCMC-MLE sampler of \citet{Hunter2006}. This estimator uses MCMC to approximate the intractable normalizing constant. However due to the iterative nature of the \citet{Hunter2006} estimator, it is less than ideal to implement in an expectation maximization framework, which in of itself is iterative. Additionally, the MCMC sampler does not scale well to larger networks, limiting the practical size of a network to less than 300 nodes. An a computationally more attractive alternative to the ML estimator appears in the maximum pseudolikelihood approach of \citet{StraussIkeda1990}. 

\subsection{Maximum Pseudolikelihood Estimation of ERGMs}

 The {\it pseudo-likelihood} approximation to the log-likelihood as developed by \citet{StraussIkeda1990} for an ERGM is then the likelihood for a binary logistic regression of the following form
\begin{equation}
P(G = g |\mathbf{X} ,\boldsymbol{\theta}) \approx \prod_{(i,j) \in \vec{A}}\frac{(\exp(\langle\boldsymbol{\theta}, \vec{s}_{ij}\rangle))^{a_{ij}}}{1+\exp(\langle\boldsymbol{\theta}, \vec{s}_{ij}\rangle)}
\end{equation}
where $\vec{s}_{ij}$ is the change in the sufficient statistics when $a_{ij}$ goes from 0 to 1, and $(i,j) \in \mathbf{A}$ refers to all $i,j$ pairs of nodes. This expression is identical to a logistic regression model treating the $\vec{s}_{ij}$ as fixed and known covariates. Taking the log of the above transforms it to the pseudo-log-likelihood which can be simplified as shown

\begin{align}
\hat{\ell}(\boldsymbol{\theta}|\mathbf{A}, \mathbf{X}) =& \log\left[ \prod_{(i,j) \in \vec{A}}\frac{(\exp(\langle\boldsymbol{\theta}, \vec{s}_{ij}\rangle))^{a_{ij}}}{1+\exp(\langle\boldsymbol{\theta}, \vec{s}_{ij}\rangle)} \right] \\
&= \log\left[ \frac{\exp(\sum_{(i,j):a_{ij} = 1} \langle\boldsymbol{\theta}, \vec{s}_{ij}\rangle)}{ \prod_{(i,j) \in \vec{A}} 1+\exp(\langle\boldsymbol{\theta}, \vec{s}_{ij}\rangle)} \right] \\
&=   \sum_{(i,j) : a_{ij} = 1} \langle\boldsymbol{\theta}, \vec{s}_{ij}\rangle- \sum_{(i,j) \in \vec{A}} \log( 1+\exp(\langle\boldsymbol{\theta}, \vec{s}_{ij}\rangle).
\end{align}

Note that $\langle\boldsymbol{\theta}, s(\mathbf{a}, \mathbf{X})\rangle$ is equal to $\sum^{(k)} \theta^{(k)} s^{(k)}(\mathbf{a}, \mathbf{X})$, and that $s^{(k)}(\mathbf{a}, \mathbf{X})$ is equal to $\sum_{(i,j) : a_{ij} = 1} s^{(k)}_{ij}$. Therefore $\sum_{(i,j) : a_{ij} = 1} \langle\boldsymbol{\theta}, \vec{s}_{ij}\rangle = \langle\boldsymbol{\theta}, s(\mathbf{a}, \mathbf{X})\rangle$. This leads to the final expression of the pseudo-log-likelihood below

\begin{equation}
\hat{\ell}(\boldsymbol{\theta}|\mathbf{A}, \mathbf{X}) = \sum_{(i,j) \in \vec{A}} \log(P(a_{ij} = 1 | \vec{A}^c_{ij})) = \langle\boldsymbol{\theta},s(\mathbf{A}, \mathbf{X})\rangle - \sum_{(i,j) \in \vec{A}} \log(1+\exp(\langle\boldsymbol{ \theta}, \vec{s}_{ij}\rangle)).
\end{equation}

$\mathbf{A}^c_{ij}$ denotes everything in $\mathbf{A}$ but $a_{ij}$. Note that this pseudo-log-likelihood differs from the log-likelihood only in the normalizing constant. Furthermore, this likelihood is identical to that of a binary logistic regression, and can be maximized by established means. 
There are several issues with using the MPLE for ERGMs that model transitivity. \cite{Lubbers2007} provide evidence that the standard errors of the estimates in an MPL fit ERGM tend to be underestimated, and the whole model tends to overfit to the data. \cite{duijna2009} show in a simulation study that the estimates of the transitivity parameter from an MPL fit ERGM are substantially less efficient than that of a ML fit ERGM. However, recently it has been proposed that the properties of the MPL estimator have not been thoroughly studied \citep{chatterjeeestimating2013} and  there is evidence for its asymptotic consistency when applied to random Markov fields with spatial structures \citep{Comets1998}, a modelling framework related to ERGMs. Furthermore, the simulation study in \citet{duijna2009} suggest that while MPLE estimates of transitivity parameters are less effecient than ML, the estimates of various covariate effects appears to be as efficient as that of the ML fit ERGMs, suggesting that the MPL estimator is appropriate to use when the interest is in the effects of various node level covariates.

With the limitations of MPLE discussed, there are several advantages that suggest MPLE is a better choice than MLE for the current purposes. Using the MLE during the maximization step is computationally intensive, with running times an order of magnitude above the running time of MPLE. This decrease in computation time is a major advantage of the MPLE approach, allowing it to scale to very large networks that would be computationally impossible to analyze using the MLE approach.  Furthermore, in light of the possibility of degeneration, repeated applications of the MLE while using non-optimal latent class solutions increases the probability that at least one iteration will be degenerate (in that, within a EM iteration, fitting the model using MLE would result in a degenerate solution from which the estimator would not be able to recover). This is not to say that however, that the MPLE approach solves the issue of degeneracy as degeneracy is a property of the model itself rather than any estimation procedure \citep{Handcock2003}. Instead, as covariate effects appear to be consistently estimated with MPLE \citep{duijna2009}, the MPLE estimator allows researchers whose main question of interest lies in covariate effects to be less concerned with potential degeneracy. However, if researchers are interested in network effects themselves, then the MPLE estimator is not appropriate, as it might mask issues with degeneracy, and lead to spurious findings. It is worthwhile to note that both of these issues (computation time, degeneracy) were raised in \citet{Steinley2010}, and motivated the use of the MPLE there as well. As such, in current manuscript, we use MPLE during the maximization step of an EM algorithm implementation. Given the known issues with MPLE, we use MLE to obtain a final set of parameter estimates, once the EM algorithm has converged. Again, this final estimation with MLE is not scalable for very large networks, while using just the MPLE approach is.

\section{Sender/Receiver Finite Mixture ERGMs}

The SRFM-ERGM developed in this manuscript is a class of models posits that every node has a latent class assignment, and this class assignment moderates the effect of specific nodal covariates on the probability of edges directed towards the node (A Receiver Finite Mixture ERGM), or edges directed away from the node (A Sender Finite Mixture ERGM), but not both. This division often aligns with hypotheses made by researchers who are interested in associations between nodal attributes and nominations. The choice to restrict the model to either assess the effect of either sender latent class or receiver latent class was made to render the model tractable to estimation via a classification likelihood EM-type algorithm \citep{kearnsan1998,macqueensome1967,Dempster1977} for reasons elaborated below.

\citet{wedela1995} developed a mixture model for generalized linear models. This approach relies on the EM type algorithm for estimation and provides an elegant solution to the problem. As the MPL for an exponential random graph model is identical to the likelihood for a binary logistic regression,  \cite{wedela1995} approach can be applied. To develop the likelihood function for the SRFM-ERGM, we use a data augmentation approach and introduce a nodal level covariate, $z_{iq}$ which is equal to $1$ if the $i$th node is a member of the $q$th  class and $0$ otherwise. $z$ is assumed to be distributed as a multinomial random variable, with $Q$ governing parameters $\boldsymbol{\pi}$. $\boldsymbol{Z}$ is an $N \times Q$ matrix of indicators of class membership.

The conditional probability density function for a network $g$ with observed adjacency matrix $\mathbf{A}$ given $\mathbf{Z}$,  $\boldsymbol{\theta}$ and $\mathbf{X}$ is

\begin{equation}
P(G = g|\mathbf{Z}, \boldsymbol{\theta}, \mathbf{X}) = \frac{\exp(\sum_{i=1}^N \sum_{q=1}^Qz_{iq}\langle \boldsymbol{\theta}_q,\mathbf{s}_{i\cdot}\rangle)}{\psi(\boldsymbol{\theta}, \mathbf{Z}, \mathbf{X})}
\end{equation}

where $\boldsymbol{\theta}_q$ is a vector of the $q$ths class's  parameters and $s_{i\cdot}$ is the contribution of node $i$'s incoming or outgoing edges (depending on if it is a receiver or sender model) to the sufficient statistics vector. $\psi(\boldsymbol{\theta}, \mathbf{Z})$ is the normalizing constant computed with the class specific parameters and given class memberships of all nodes. When $\mathbf{Z}$ is unobserved the conditional probability density function for $\mathbf{Z}$ is

\begin{equation}
    P(\mathbf{Z}|\boldsymbol{\pi}) = \prod_{i= 1}^N \prod_{q=1}^Q \pi_q^{z_{iq}}.
\end{equation}

Therefore the complete-data likelihood for an SRFM-ERGM is

\begin{equation}
    L(\boldsymbol{\theta}, \boldsymbol{\pi}, \mathbf{Z}|\mathbf{X},\mathbf{A}) =  \frac{\exp(\sum_{i=1}^N \sum_{q=1}^Q z_{iq}\langle \boldsymbol{\theta}_q,\vec{s}_{i\cdot}\rangle)}{\psi(\boldsymbol{\theta},\mathbf{Z})} \prod_{i= 1}^N \prod_{q=1}^Q \pi_q^{z_{iq}}
\end{equation}

and the complete-data log likelihood is 
\begin{equation}
    \ell(\boldsymbol{\theta}, \boldsymbol{\pi},\mathbf{Z}|\mathbf{X},\mathbf{A}) =  \sum_{i=1}^N \sum_{q=1}^Q z_{iq}\langle \boldsymbol{\theta}_q,\vec{s}_{i\cdot}\rangle- \log(\psi(\boldsymbol{\theta}, \mathbf{Z})) + \sum_{i= 1}^N \sum_{q=1}^Q z_{iq} \log(\pi_q).
\end{equation}

While this complete-data log likelihood allows for the direct calculation of the expected class membership probabilities $\mathbf{E}[\mathbf{Z}| \boldsymbol{\theta}]$ once $\boldsymbol{\theta}$ is known, due to the presence of the normalizing constant $\psi(\boldsymbol{\theta}, \mathbf{z})$ there are still computational difficulties in estimating $\boldsymbol{\theta}$. As such, we use the log pseudo-likelihood. 

The complete-data log pseudo-likelihood for a SRFM-ERGM is

\begin{equation}
\hat{\ell}(\boldsymbol{\hat{\theta}}, \boldsymbol{\pi}, \mathbf{Z}|\mathbf{A}, \mathbf{X}) = \sum_{i} \sum_{q=1}^{Q} z_{iq}\log(\pi_q) +  \sum_{i} \sum_{j \neq i}\left[ a_{ij} \left[ \sum_{q=1}^{Q} z_{iq}   \langle\boldsymbol{\theta}_{q},\vec{s}_{ij} \rangle \right] - \sum_{q=1}^{Q} z_{iq}\log(1+\exp(\langle\boldsymbol{\theta}_{q} ,\vec{s}_{ij}\rangle))\right]
\end{equation}

 where $\pi_q$ is the marginal probability of a node being in class $q$. $\boldsymbol{\theta}_q$ is a vector of parameters for class $q$. $a_{ij}$ is the value (0 or 1 for a binary network) of the edge from node $i$ to node $j$. $s_{ij}$ is the vector of changes in the sufficient statistics of the network if edge $a_{ij}$ went from 0 to 1, given the rest of the network. Any given parameter in $\boldsymbol{\theta}$ may be restricted to be equal across all classes and we term these common parameters as \textit{homogeneous}. 

The choice to restrict these models to mixture components defined by either sender or receiver nodes is due to the dependency structure noted by \citet{Daudin2008} in their development of mixture models for Bernoulli networks. Should the mixture component be defined by both incoming and outgoing ties then the vector $\theta_q$ would actually be $\theta_{ql}$ where $l$ is the mixture component label of the receiving node. As such, each edge in the network would have a likelihood that is dependent on the mixture component of both the sending and receiving node. In turn, as was noted by \citet{Daudin2008}, the expected value of a mixture component label for individual $i$ becomes dependent on the component label of every other node in the network. This requires that the expected value of component label be estimated simultaneously for every node, leading to a sum over every possible combination of component labels in the denominator, which quickly becomes computationally infeasible. The choice to restrict the models to either sender or receiver leads to a separable expected value statement for each node.

\subsection{Expectation Step}

The general expectation for a latent class general linear model (where the probability densities are exponential family) is in the following form \citep[Eq 7;][]{wedela1995}

\begin{equation}
E[z_{iq} | \boldsymbol{\theta},\boldsymbol{\pi}, \vec{y}_i] = \frac{\pi_q \prod^K_{k=1}f_{ik|q}(\vec{y}_{ik}|\vec{\theta_q})}{\sum_{q=1}^{Q} \pi_q \prod^K_{k=1}f_{ik|q}(\vec{y}_{ik}|\vec{\theta_q})}
\label{desarb}
\end{equation}

where $\vec{y}_i$ is a vector of observations (each of which can come from a different arbitrary probability density) for case $i$ and $f_{ik|q}(\cdot)$ is the probability density function for the $k$th variable of the $i$th case given the $i$th case is in class $q$.

 Here we can express the expectation for class membership of a given node in precisely that form of Equation \ref{desarb}

\begin{equation}
E[z_{iq} | \boldsymbol{\theta}, \boldsymbol{\pi}, \vec{A}, \mathbf{X}] = \frac{\pi_q \exp ( \sum_{j \neq i} \left[ a_{ij} \left[    \langle\boldsymbol{\theta}_{q},\vec{s}_{ij}\rangle \right] -  \log(1+\exp(\langle\boldsymbol{\theta}_{q},\vec{s}_{ij}\rangle))\right])}{\sum_{q=1}^{Q} \pi_q  \exp( \sum_{j \neq i} \left[ a_{ij} \left[    \langle\boldsymbol{\theta}_{q},\vec{s}_{ij}\rangle \right] - \log(1+\exp(\langle\boldsymbol{\theta}_{q},\vec{s}_{ij}\rangle))\right]) }.
\end{equation}

Note that in the above likelihood, edges that share the same sender node all are assigned the same latent class. Intuitively, this is due to the assignment of the latent class being sender based rather than edge based. This maps onto the Wedel and DeSarbo approach, as they classify observation vectors. In this case the observation vector is the vector of all possible edges a node could and did send. 

With the expectation derived, we diverge from the approach laid out in \citet{wedela1995}. In \citet{wedela1995}'s approach, the expectations produce probabilities of class membership, and these probabilities are then used in the maximization step. This approach is known as mixture EM, which is the EM algorithm laid out originally by Dempster, Rubin and Laird \citeyear{Dempster1977}. In this manuscript, we take a classification likelihood approach. The classification likelihood approach (for examples see: \cite{Symons1981,McLachlan2005,Steinley2011a}) assigns each case the class label that has the highest probability. This hard assignment of classes implies that although the likelihood of the data will be a mixture across the classes, each case is not a mixture across the latent classes. Classification likelihood can be fit using an EM-type algorithm that uses the class label with the highest probability instead of using the expected value of class membership (a mixture approach).

This hard assignment has advantages and disadvantages. Classification likelihood algorithms consistently find more informative latent classes than mixture algorithms in the sense that the Kullback-Liebler  divergence \citep{Kullback1951} of the mixture components is maximized; however it does not maximize the likelihood of the data as the mixture EM algorithm does \citep{kearnsan1998}.  In terms of disadvantages, classification likelihood does not recover ill-separated mixture components well, nor does it recover very unbalanced group sizes \citep{Celeux1993,Govaert1996}. This failure to recover ill-separated groups comes from the maximization of the KL divergence. If the groups are ill-separated, classification likelihood tends to model the data using a single distribution in which the majority of cases are placed. This can be an advantage if the presence of latent classes is being tested; however, if the presence of latent classes is assumed, this approach will not detect them if they are ill-separated. This makes classification likelihood more conservative in its detection of latent classes, whereas mixture likelihood approaches could result in the over extraction of the latent classes \citep{Bauer2003b}. Additionally classification likelihood does not exhibit optimal large sample properties in that it tends to be asympototically biased \citep{Bryant1991}, though this bias appears to be lessened when allowing for different cluster sizes \citep{Celeux1993}, which is what the estimation for the SRFM-ERGM allows. However, this property of the classification likelihood makes it more suitable for the clustering of observations than a traditional mixture-likelihood approach \citep{Steinley2011}. The finding of more informative clusters, as well as classification likelihoods tendency to be more conservative in its clustering are two major advantages of this approach. By using classification likelihood, the Sender/Receiver Finite Mixture ERGMs are explicitly attempting to classify individuals into well separated clusters, rather than approximate a complex distribution using mixtures of simpler distributions.

\subsection{Maximization Step}

With the latent class labels assigned, the conditional likelihood is
\begin{equation}
\hat{\ell}(\boldsymbol{\theta}, \boldsymbol{\pi} | \mathbf{Z}, \vec{A}, \mathbf{X}) = \sum_{i} \sum_{q=1}^{Q} z_{iq}\log(\pi_q) + \sum_{(i,j) \in \vec{A}} \left[  a_{ij} \left[ \sum_{q} Z_{iq} \langle\boldsymbol{\theta}_{q},\vec{s}_{ij}\rangle \right] - \sum_{q} Z_{iq}\log(1+\exp(\langle\boldsymbol{\theta}_{q},\vec{s}_{ij}\rangle )) \right],
\end{equation}

where $z_{iq}$ is a binary indicator of node $i$'s membership in latent class $q$. Note here that component of the likelihood involving $\boldsymbol{\pi}$ is separate from the component involving $\boldsymbol{\theta}$. With $\mathbf{Z}$ given, the maximum likelihood estimate of $\boldsymbol{\pi}$ is simply the proportion nodes in each class. The remaining component of the likelihood is the same as the likelihood for a binary logistic regression, and as such can be maximized in $\boldsymbol{\theta}$ using standard estimation methods for logistic regression \citep{Nelder1972}.

\subsection{A Note on Standard Errors}

In Wedel and DeSarbo's \citeyear{wedela1995} description of a mixture model on the general linear model, they note that the standard errors of the estimates can be calculated using a weighted information matrix, taking into account the uncertainty of latent class assignments. However, this relies on the mixture EM approach. In a classification EM-like approach, the parameters are estimated at each maximization step with each node assigned to a class with a weight of 1. As such, the standard errors of the estimates can be interpreted as the standard error of the estimated parameters if the estimated latent class solution was observed and true. This does neglect the uncertainty regarding the latent class labels, and as such, the standard errors of the estimates are likely underestimated.

\subsection{Estimation for SRFM-ERGMs}

There are two additional considerations for estimating SRFM-ERGMs. The first is that of multiple start values. The EM algorithm, either classification or mixture type, is susceptible to local maxima \citep{hipplocal2006,Dempster1977}. To account for this, we initialize the estimation with multiple start values. Once every model has converged, the latent class solution with the greatest likelihood is selected.

Additionally, there are the multiple issues with using the MPLE to obtain parameter estimates as previously described. To account for this, once the EM algorithm has converged the estimated latent class labels are then used as observed class labels to fit an ERGM model using MLE. The parameter estimates of that model are presented.

The full estimation algorithm proceeds as follows:

\begin{enumerate}
\item Initialize starting values for latent class labels $\vec{Z}_0$
\item For iteration r\begin{enumerate}
\item Maximize the likelihood $\ell(\boldsymbol{\theta}_r|\mathbf{Z}_{r-1})$ to obtain $\hat{\boldsymbol{\theta}}_r$ and $\hat{\boldsymbol{\pi}_r}$
\item Compare  $\ell(\boldsymbol{\theta}_r|\mathbf{Z}_{r-1})$ against $\ell(\boldsymbol{\theta}_{r-1}|\mathbf{Z}_{r-2})$.  
\item Obtain the hard latent class labels $\mathbf{Z}_r$ using $E(\mathbf{Z}_r|\boldsymbol{\theta}_r)$ If change in log-likelihood is less than tolerance, estimation has converged to a solution, move to Step 3.
\item Return to Step 1
\end{enumerate}
\item Once estimation has converged, save solution, initialize a new set of starting values and return to step 2.
\item Once solutions are saved for a pre-defined number of starting values, choose the converged solution with the greatest final $\ell(\boldsymbol{\theta}|\mathbf{Z})$. This solution has the most probable set of latent class labels out of all of the solutions.
\item With the most probable solution selected, fit a MLE ERGM considering the latent class labels as observed, and return the parameter estimates $\boldsymbol{\hat{\theta}}$.
\end{enumerate}

\section{Empirical Example: Heterogeneous Peer Assortment of Alcohol Use}

To motivate the use of the SRFM-ERGM approach, in this empirical example we examine potential heterogeneity in the effect of alcohol use on peer assortment. Peer assortment on the basis of substance use and other deviant behaviors has been a continually studied topic in clinical psychology \cite{brechwaldbeyond2011}, however whenever network models are applied to network data to study peer assortment on substance use, researchers have used models that assume all individuals are governed by the exact same set of parameters, in other words that the effects of all covariates are \textit{homogeneous}. Heterogeneous effects of alcohol use on peer assortment could potentially have important implications for network intervention design, as an intervention model based on homogeneous results may not be applicable to all at risk individuals in the intervention population. 

We examine alcohol use here due to its fairly high use rate among this sample (26.67\% of the sample have used alcohol). Additionally, we examine the effects of marijuana use, tobacco use and anti-social behaviors on peer assortment.

Three parameters were used to model each of the effects of alcohol use, marijuana use, tobacco use and anti-social behavior. Indegree parameters measure the effect of the variable on an individuals tendency to receive friendship nominations, outdegree parameters measure the effect of a variable on an individuals tendency to send friendship nominations. Finally, absolute difference terms measure the effect of different values of the variable on the tendency for two individuals to form a friendship.

In the heterogeneous model, we propose a sender-type heterogeneity for the outdegree effect of alcohol use and the absolute difference effect of alcohol use. This can be interpreted as allowing the effect of alcohol use to differ in its effect on an individual's tendency to nominate friends both generally, and with regards to the potential friends' alcohol use. 

\section{Data Characteristics and Measures}

This empirical data-set consisted of a 151 students who attended the same high school. Students were assessed on a variety of demographics and risk behaviors. The current analysis focuses on substance use of three types, alcohol, tobacco and marijuana, as well as general anti-social behavior. Descriptive statistics of the data are provided in Table \ref{demo}. 

\begin{table}[H]
\centering
\caption{Demographic and Descriptive Statistics}
\label{demo}
\resizebox{\textwidth}{!}{%
\begin{tabular}{lllllll}
\hline
                     & Male             & Female        & Missing         &               &            &         \\ \hline
Gender               & 46\%             & 51.30\%       & 2.60\%          &               &            &         \\ \hline
                     & African-American & Asian         & Latino/Hispanic & White         & Mixed      & Missing \\ 
Ethnicity            & 23.03\%          & 1.32\%        & 18.42\%         & 48.03\%       & 6.58\%     & 2.63\%  \\ \hline
                     & Mean             & 25\% Quantile & Median          & 75\% Quantile & \% Above 1 & Missing \\ 
Age                  & 15.07            & 14.65         & 15.01           & 15.33         & -          & 0.66\%  \\
Alcohol Use          & 0.4476           & 0             & 0               & 1             & 26.97\%    & 5.92\%  \\
Tobacco Use          & 0.1724           & 0             & 0               & 0             & 6.58\%     & 4.61\%  \\
Marijuana Use        & 0.1781           & 0             & 0               & 0             & 9.21\%     & 3.95\%  \\
Anti social behavior & 0.305            & 0.129         & 0.258           & 0.4516        & -          & 4.61\%  \\ \hline
\end{tabular}
}
\end{table}

\subsection{Measures}

{\it Friendship Nominations.} Individuals were provided a full roster of the school and were allowed to nominate any number of individuals as friends. These nominations were binary (Yes; a friend, or No; not a friend), and directed.

{\it Alcohol Use.} Alcohol use was assessed with a single item which asked "In the last 6 months on how many days did you have at least one drink of alcohol?" With the following answer choices: 1) 0 days, 2) 1-2 days, 3) 3-5 days, 4) 6-9 days and 5) 10 or more days.

{\it Tobacco Use.} Tobacco use was assessed with a single item which asked "In the last 6 months on how many cigarettes did you smoke per day?" With the following answer choices: 1) 0, 2) 1, 3) 2-3 , 4) 6-10 and 5) 10-20, 6) more than a pack.

{\it Marijuana Use.} Marijuana use was assessed with a single item which asked "In the last 6 months how many times did you use marijuana?" With the following answer choices: 1) 0, 2) 1-2 times, 3) 3-9 times, 4) 10-19 and 5) 20 or more times.

{\it Anti Social Behavior.} Antisocial behavior was assessed with a subset of 31 items from the Youth Self Report scale \citep{achenbachmanual1991}. These items assessed behaviors over the previous 6 month period and had text such as "I get in many fights", "I set fires," and "I tease others a lot." Responses were on a 3 point scale with 0 being "Not True", 1 being "Somewhat True", and 2 being "Very/Often True." These items were mean scored for a final Anti-Social behavior Composite.

\subsection{Missing Data}

Due to the inability for ERGMs to handle missing data at the nodal attribute level, multiple imputation was used to account for missingness \citep{rubinmultiple1987}. 500 covariate datasets were simulated using information on gender, age, ethnicity, substance use and all anti social behavior items. Results for the 500 covariate datasets were combined according to \citet{rubinmultiple1987}. There are two competing interests to the use of multiple imputation here. As was mentioned, ERGMs cannot account for missingness among the nodal covariates, as such one option would be to remove those individuals from the network. However, as networks are highly interdependent systems by definition, this would substantially change the network structure, possibly leading to different results. Multiple imputation however, does not take into a account the dependency between individuals on their nodal attributes, rather assuming them to be independent. This assumption leads to an attenuation of any homophily effects within the network. In the case of this application, it was decided that the damage to the network structure outweighed the attenuation of homophily, and multiple imputation was used. 

\subsection{Model Specification}

In this example, we are specifically interested in \textit{sender} effects of alcohol use and sender effects of differences in alcohol use. Substantively, these effects can be interpreted as the association between alcohol use and raw number of friendship nominations made, and the impact of the difference in alcohol use on the probability of a friendship nomination being made. To demonstrate the utility of the SRFM-ERGM approach, we fit a 2 class model with heterogeneous sender effects of alcohol use, and sender effects of the absolute difference in alcohol use. Additionally, we allowed the edge parameter to vary between the two classes. We term the model with these heterogeneous effects as the \textit{heterogeneous model}.  Furthermore, we fit a model with no heterogeneous effects so as to compare parameter estimates. This model we term the \textit{homogeneous model}. A detailed description of all effects estimated follows:

\subsection{Structural Statistics}

    {\it Edges.} The parameter estimate associated with the edges statistic in any ERGM acts as an intercept for the model. The sufficient statistic described by this term is a count of the number of edges in the network. In the homogeneous model, a single edge parameter was estimated. In the heterogeneous model, each class had its own edge parameter.
    
    {\it Mutuality.} The sufficient statistic is the total number of reciprocated dyads (Holland \& Leinhart, 1981). These are dyads where $a_{ij} = a_{ji} = 1$. The parameter associated with this statistic can be described as the tendency for edges to be reciprocated.
    
    {\it Geometrically Weighted Edgewise Shared Partners (GWESP).} This sufficient statistic is used to account for transitivity in the network \citep{Hunter2008}. Transitivity is the phenomena where an edge is more likely to be present if the edge has shared partners. This is commonly referred to colloquially as ``A friend of a friend is my friend.'' Transitivity is a common phenomena in real-world networks, and failure to model transitivity can result in bias in the estimates of other effects \cite{duijna2009}. However, when a term that models transitivity is added, this induces a dependency structure in the network that prevents a direct maximization of the likelihood during estimation. This dependency structure also runs the risk of causing degeneracy in estimation, where the expected value of the network's distribution converges to a completely empty or completely full network. Details of why this occurs can be found in \citet{chatterjeeestimating2013}. The geometrically weighted edgewise shared partners term works to prevent degeneracy by down-weighting large numbers of shared partners. A GWESP term has both an effect parameter $\theta^{(GWESP)}$ and a weight parameter $\tau$. $\tau$ ranges from 0 to $\infty$, with 0 meaning that the effect of shared partners does not increase beyond having 1 shared partner, and $\infty$ meaning that the effect of additional shared partners is linearly increasing with sample size.
    
    If $\tau$ is allowed to be freely estimated, then the ERGM becomes a member of a curved exponential family \citep{efronthe1978}, which complicates estimation. In the simulations that follow, $\tau$ is considered fixed and known, a practice that is done in substantive analysis where the curved exponential family models fail to converge \citep{Hunter2008}. In this case we restrict $\tau = .1$, representing a sharp drop off in the effect of shared partners as the number of shared parents increase. The empirical results were robust within a range of $\tau$, between 0 and .25.
    
\subsection{Nodal Attribute Terms}
    
    {\it Attribute Matching.} The parameter estimate associated with this sufficient statistic is used to evaluate the effect of nodes matching on a categorical variable (such as gender or ethnicity) on the probability of having an edge . The sufficient statistic is the number of edges between nodes that match on the categorical variable. Gender and ethnicity matching are modeled in both the heterogenous and homogeneous models.
    
    {\it Sender Attribute} The parameter estimate associated with this sufficient statistic is used to evaluate the interaction effect of a continuous variable and the number of edges coming out of nodes on the probability of having an edge. The node level sufficient statistic is the outdegree of a node multiplied by the attribute value for that node. Tobacco use, anti-social behavior and marijuana use all have a homogeneous sender effect in both models, while alcohol use has a homogeneous sender effect in the homogeneous model, and a 2-class heterogeneous sender effect in the heterogeneous model. 
    
    {\it Receiver Attribute} Similarly to the Sender Attribute term, the parameter estimate associated with this sufficient statistic reflects how a continuous variable differentially relates to the probability of an edge based on the number of edges the node receives. The node level sufficient statistic is indegree of the node multiplied by the attribute value for that node. Tobacco use, anti-social behavior, marijuana use and alcohol use have a homogeneous receiver effect in both models.
    
    {\it Attribute Absolute Difference} The parameter associated with this sufficient statistic evaluates the effect of similarity on a continuous variable between nodes on the probability of an edge. The edge-level sufficient statistic is the absolute difference in the values of the attribute for the sender and receiver of the edge multiplied by the presence or absence of that edge. Tobacco use, anti-social behavior and marijuana use all have a homogeneous absolute difference effect in both models, while alcohol use has a homogeneous absolute difference effect in the homogeneous model, and a 2-class heterogeneous sender absolute difference effect in the heterogeneous model. This sender specific effect models the probability that an individual will nominate another individual with differing alcohol use.

\section{Results of the Empirical Analyses}

\begin{table}[H]
\centering
\caption{Empirical Results for both heterogeneous model and homogeneous model. Estimates are in mean log-odds metric calculated across 500 multiply imputed datasets with (multiple imputation standard error)}
\label{empres}
\singlespacing
\begin{tabular}{lrr}
\hline
Parameter                          & Heterogeneous Model & Homogeneous Model \\ \hline
GWESP ($\tau$ = .1)                 & 0.97 (0.07)*        & 1.25 (0.07)*      \\
Mutuality                          & 2.34 (0.14)*        & 2.26 (0.15)*      \\
Gender Match                       & 0.52 (0.06)*        & 0.44 (0.06)*      \\
Ethnicity Match                    & 0.85 (0.06)*        & 0.47 (0.06)*      \\
Sender Tobacco Use                 & 0.13 (0.11)         & 0.29 (0.10)*      \\
Receiver Tobacco Use               & -0.11 (0.10)        & -0.05 (0.09)      \\
Sender Anti Social                         & -0.64 (0.21)*       & -0.23 (0.20)      \\
Receiver Anti Social                       & 0.24 (0.20)         & 0.24 (0.20)       \\
Sender MJ                          & 0.33 (0.11)*        & 0.14 (0.10)       \\
Receiver MJ Use                    & 0.21 (0.11)         & 0.19 (0.10)       \\
Receiver Alcohol Use               & 0.04 (0.07)         & -0.02 (0.06)      \\
Abs Difference Anti Social                 & -0.08 (0.11)        & -0.02 (0.18)      \\
Abs Difference Tobacco Use         & -0.21 (0.10)*       & -0.13 (0.10)      \\
Abs Difference MJ Use              & -0.39 (0.19)*       & -0.18 (0.10)      \\
Abs Difference Alcohol Use Class 1 & -0.01 (0.08)        & -0.01 (0.06)      \\
Abs Difference Alcohol Use Class 2 & -0.05 (0.07)        &                   \\
Sender Alcohol Use Class 1         & -0.21 (0.10)*       & 0.02 (0.06)       \\
Sender Alcohol Use Class 2         & 0.06 (0.07)         &                   \\
Edges Class 1                  & -5.28 (0.11)*       & -5.09 (0.10)*     \\
Edges Class 2                  & -4.22 (0.12)*       &                   \\ \hline
\end{tabular}

\begin{tablenotes}
\item MJ: Marijuana, Abs Difference: Absolute Difference
\item *: p \textless .05
\end{tablenotes}
\end{table}

Table \ref{empres} contains the results for the heterogeneous and homogeneous analysis. Of note in these results is the effect of Sender Alcohol Use in Class 1  versus the effect of Alcohol Use in Class 2 . These effects are significantly different from one another (z = -2.21, p \textless .05), suggesting that for members of Class 1, alcohol use reduces the number of friendship nominations sent out while for members of Class 2, alcohol use does not effect the number of friendship nominations sent out. Additionally, there is a significant difference between the edges effect of Class 1 and Class 2 (z = 6.51, p \textless .05). This suggests that members of Class 1 have fewer friendship nominations made in general, while members of Class 2 make more friendship nominations. 

There are several significant homogeneous effects in the heterogeneous model. The effect of Sender Anti Social Behavior is negative suggesting that individuals with more antisocial behavior nominate fewer friends. The effects of Absolute Difference in Tobacco and Marijuana Use are both negative suggesting that individuals who are different on their usage of those substances are less likely to nominate each other as friends. Finally there was a positive significant effect of Sender Marijuana Use  suggesting that individuals who use more Marijuana make more friendship nominations in general.

There are expected positive significant findings for GWESP, mutuality, gender and ethnicity matching. The significant GWESP finding suggests that individuals who share friends are more likely to be friends themselves. Additionally there is a strong effect for mutuality suggesting that reciprocal friendship nominations are the norm in this network. 

As for the differences between the heterogeneous model and the homogeneous model, it is apparent that the modeling of heterogeneous effects of alcohol use and the edge parameter clarifies several of the homogeneous parameters. Specifically, the effects for Sender Anti Social Behavior, Sender Marijuana Use, Absolute Difference in Tobacco Use and Absolute Difference in Marijuana Use are significant in the Heterogeneous Model and not significant in the Homogeneous Model. Interestingly, one effect, that of Sender Tobacco Use, was significant in the homogeneous model and not significant in the heterogeneous model. 

It is worthwhile to note that significantly fewer individuals were classified into class 2 than into class 1, with the average proportion of individuals in class 2 across all the multiply imputed datasets being 27.8\%. Class separation on the basis of class membership probability was exceptionally good for every multiply imputed dataset. The average probability of membership in class 1 for individuals with a higher than .5 probability of being in class 1 was .939, while the average probability of class membership in class 2 for individuals with a higher than .5 probability of being in class 2 was .9671. This level of class separation is in part due to the classification EM algorithm used in estimation, which tends to find the latent class labeling that is the most informative rather than the latent class probabilities that maximize the likelihood. 

In summary, the findings support a level of heterogeneity in the effect of alcohol use, specifically for the sender effect of alcohol use. Indeed, some individuals are less likely to nominate friends the more alcohol they drink, while other individuals exhibit no effect of alcohol use on friendship formation. Of broader interest, there are more significant findings in the heterogeneous model as compared to the homogeneous model. This suggests that the inclusion of the heterogeneous model parameters clarified the effects of the other parameters.

To better explore the consequences of failing to model heterogeneous effects, and to evaluate the performance of the SRFM-ERGM framework, we now present a set of simulation studies based on the previous empirical example. The necessary code for fitting SRFM-ERGM models is contained in the Supplementary Materials, and users interested in the simulation study data and generating scripts can contact the lead author for access.

\section{Simulation Study}

In order to test the performance of the SRFM-ERGM and to assess the consequences of neglecting to model latent classes, we simulated networks from 9 simulation conditions, and examined the simulated networks in three sets of analyses. Analysis 1 demonstrated the consequences of fitting homogeneous ERGM models to networks that are truly heterogeneous. The Analysis 2 demonstrated the improvement in recovery of model parameters when using SRFM-ERGMs on heterogeneous networks. Finally, the Analysis 3 demonstrated improvement in latent class labels as sample size increases. 

\subsection{Simulation Strategy}

To more accurately reflect the type of data a researcher would encounter as well as to follow good practice in ERGM methodology development \citep{duijna2009}, the simulations were based off of the previous empirical dataset.  Heterogeneity on sender alcohol use, sender absolute difference in alcohol use, and the edge parameter was modeled in the empirical analysis.

Due to missing data at the covariate levels, multiple imputation was used to generate a set of covariate datasets for the empirical analysis. In these simulations, a single dataset of covariates was randomly selected from the multiply imputed datasets, and was used for all simulated networks. This follows the simulation procedure outlined in \citet{duijna2009}. For each simulation trial, a new network was simulated without changing the covariate dataset.

There were 9 simulation conditions in total (see Table \ref{SimTable}). Simulation conditions were arrived at by examining the parameter estimates from the empirical model and changing some of the parameter values to test how different latent class definitions impacted inference. The conditions were as follows:

\begin{itemize}
\item \textit{Condition 1: Heterogeneous Edge Parameter.} This condition specified that the two latent classes differed only in the value of their edge parameter, so that one class had, on average, a lower number of edges sent than the other class.
\item \textit{Condition 2: Heterogeneous Covariate Parameter. } This condition specified that the two latent classes differed only on the value of their Sender Alcohol Use effect, so that in Class 1 Alcohol Use had no effect on the number of sent edges, while in Class 2, increased Alcohol Use decreased the number of sent edges.
\item \textit{Condition 3: Heterogeneous Edge and Covariate Parameter.} This condition specified that the two latent classes differed in both their Edge effect and their Sender Alcohol Use effect. Class 1 had no effect of Alcohol Use, as well as on average higher numbers of sent edges, while Class 2 had increased Alcohol Use leading to decreased number of sent edges, as well as lower number of sent edges overall.
\item \textit{Condition 4: Heterogenous GWESP Parameter.} In this condition the latent classes differed only in the value of their GWESP parameter. Class 1 had a higher value of the GWESP parameter than Class 2, corresponding to greater rates of transitive closure.
\item \textit{Condition 5: Homogeneous Model} In this condition there were no latent classes present.
\end{itemize}

 Additionally, for all conditions the edge parameter was increased from the empirical example to simulate less sparse matrices and therefore increase the amount of information in the data. Finally, in all conditions, approximately 25\% of nodes were in class 2. This was derived from the empirical results. Additionally, we examine the effect of increasing class separation by increaing the effect size of heterogeneous parameters in a set of additional conditions. These increased effect size conditions are denoted with a +.

\begin{table}[H]
\centering
\caption{Simulation Conditions: Values are the data generating parameters per condition.}
\label{SimTable}
\resizebox{\textwidth}{!}{%
\begin{tabular}{lrrrrrrrrr}
\hline
Condition                  & \multicolumn{1}{c}{1} & \multicolumn{1}{c}{1+} & \multicolumn{1}{c}{2} & \multicolumn{1}{c}{2+} & \multicolumn{1}{c}{3} & \multicolumn{1}{c}{3+} & \multicolumn{1}{c}{4} & \multicolumn{1}{c}{4+} & \multicolumn{1}{c}{5} \\ \hline
GWESP (.1) Class 1         & 0.97                  & 0.97                   & 0.97                  & 0.97                   & 0.97                  & 0.97                   & 1                     & 1.2                    & 0.97                  \\
GWESP (.1) Class 2         & 0.97                  & 0.97                   & 0.97                  & 0.97                   & 0.97                  & 0.97                   & 0.2                   & 0.2                    & 0.97                  \\
Mutual                     & 2.34                  & 2.34                   & 2.34                  & 2.34                   & 2.34                  & 2.34                   & 2.34                  & 2.34                   & 2.34                  \\
Gender Match               & 0.52                  & 0.52                   & 0.52                  & 0.52                   & 0.52                  & 0.52                   & 0.52                  & 0.52                   & 0.52                  \\
Ethnicity Match            & 0.84                  & 0.84                   & 0.84                  & 0.84                   & 0.84                  & 0.84                   & 0.84                  & 0.84                   & 0.84                  \\
Sender Tobacco Use         & 0.13                  & 0.13                   & 0.13                  & 0.13                   & 0.13                  & 0.13                   & 0.13                  & 0.13                   & 0.13                  \\
Receiver Tobacco Use       & -0.11                 & -0.11                  & -0.11                 & -0.11                  & -0.11                 & -0.11                  & -0.11                 & -0.11                  & -0.11                 \\
Sender Anti Social                 & -0.64                 & -0.64                  & -0.64                 & -0.64                  & -0.64                 & -0.64                  & -0.64                 & -0.64                  & -0.64                 \\
Receiver Anti Social               & 0.24                  & 0.24                   & 0.24                  & 0.24                   & 0.24                  & 0.24                   & 0.24                  & 0.24                   & 0.24                  \\
Sender MJ                  & 0.33                  & 0.33                   & 0.33                  & 0.33                   & 0.33                  & 0.33                   & 0.33                  & 0.33                   & 0.33                  \\
Receiver MJ                & 0.21                  & 0.21                   & 0.21                  & 0.21                   & 0.21                  & 0.21                   & 0.21                  & 0.21                   & 0.21                  \\
Receiver Alcohol Use       & 0.04                  & 0.04                   & 0.04                  & 0.04                   & 0.04                  & 0.04                   & 0.04                  & 0.04                   & 0.04                  \\
Abs Diff Tobacco Use       & -0.21                 & -0.21                  & -0.21                 & -0.21                  & -0.21                 & -0.21                  & -0.21                 & -0.21                  & -0.21                 \\
Abs Diff MJ                & -0.39                 & -0.39                  & -0.39                 & -0.39                  & -0.39                 & -0.39                  & -0.39                 & -0.39                  & -0.39                 \\
Abs Diff Anti Social               & -0.08                 & -0.08                  & -0.08                 & -0.08                  & -0.08                 & -0.08                  & -0.08                 & -0.08                  & -0.08                 \\
Abs Diff Alcohol Use       & -0.01                 & -0.01                  & -0.01                 & -0.01                  & -0.01                 & -0.01                  & -0.01                 & -0.01                  & -0.01                 \\
Sender Alcohol Use Class 1 & 0.06                  & 0.06                   & 0                     & 0                      & 0                     & 0                      & 0.06                  & 0.06                   & 0.06                  \\
Sender Alcohol Use Class 2 & 0.06                  & 0.06                   & -0.25                 & -1                     & -0.25                 & -1                     & 0.06                  & 0.06                   & 0.06                  \\
Edges Class 1              & -3.5                  & -3.25                  & -3.85                 & -3.85                  & -3.5                  & -3.25                  & -3.65                 & -3.65                  & -3.85                 \\
Edges Class 2              & -4.25                 & -4.5                   & -3.85                 & -3.85                  & -4.25                 & -4.75                  & -3.65                 & -3.65                  & -3.85                 \\ \hline
\end{tabular}
}
\begin{tablenotes}
\item MJ: Marijuana, Abs Diff: Absolute Difference
\end{tablenotes}
\end{table}

In Conditions 1 through 4 and 1+ though 4+, correctly specified \textit{heterogeneous} models were fit to the data as well as homogeneous models. In Condition 5 a model that specified 2 latent classes and heterogeneity on the sender alcohol use parameter and the edges parameter were estimated, along with a homogeneous (the correctly specified) model. 

All conditions had 500 networks simulated. The networks were simulated using the MCMC approach used in the R package statnet \citep{statnet:2008}. This approach randomly toggles edges within a network with the probability according to the generating model. This allows the simulated data to properly reflect transitivity, as well as contain sampling variability. To clarify, only the networks themselves were simulated. A single covariate data set chosen at random from the 500 multiply imputed empirical example datasets was used across all simulations, as were the latent class labels. Estimation of these models used twenty random start values per trial as per the algorithm described above. 

Finally in our third analysis the size of the network $N$ was manipulated. In an attempt to keep the nodal covariate distributions similar as sample size increased, for each value of $N$, the node level covariate data was formed by sampling with replacement from the 151 original individuals' data. Once one dataset per value of $N$ was generated, 500 networks using Condition 3 parameter values and 500 networks using Condition 4 parameter values were generated and correctly specified heterogeneous models were fit. These conditions were chosen to examine the effect of increased sample size on classes defined by dyadic effects (Condition 3) and on transitivity based effects (Condition 4).

\subsubsection{Outcomes} Raw and relative biases were assessed, as well as computed standard errors from the simulation set and average estimated standard errors. Additionally, Hubert-Arabie Adjusted Rand Indices \citep[ARI;][]{Hubert1985c,steinleyproperties2004} were computed for the comparison of the estimated latent class labels and the true latent class labels.  The Adjusted Rand Index is a measure of agreement that accounts for the expected level of agreement if the labels were assigned randomly. Finally, for Conditions 2 and 2+ Adjusted Rand are presented for the latent class labels of only the nodes that had a value higher than 0 for Alcohol Use.  By examining the subset of nodes with positive Alcohol Use, we can examine the performance of the method when the latent classes are defined only for a subset of nodes.

\section{Results}

\subsection{Analysis 1: Homogeneous Models applied to Heterogeneous Data}

What follows are the results from fitting homogeneous models to the data generated from the conditions. Relative biases are presented in Table \ref{homorel}. Tables of mean parameter value estimates, raw bias and estimates of standard errors are contained in Appendix A. Note that for Conditions 1 through 4+, the homogeneous model fitted is not the generative model, while for Condition 5, the homogeneous model is the true model.

\begin{table}[H]
\centering
\caption{Relative Bias for Homogeneous Models calculated with Mean Raw Bias over True Parameter Values. Results indicate that there is a substantial and systematic pattern of bias in the homogeneous parameters when heterogeneity is not properly modeled. Relative bias for heterogeneous parameters reflects using the single parameter estimate as the estimate for both true class parameters.}
\label{homorel}
\resizebox{\textwidth}{!}{%
\begin{tabular}{lrrrrrrrrr}
\hline
Condition                  & \multicolumn{1}{c}{1}          & \multicolumn{1}{c}{1+}         & \multicolumn{1}{c}{2}          & \multicolumn{1}{c}{2+}         & \multicolumn{1}{c}{3}          & \multicolumn{1}{c}{3+}         & \multicolumn{1}{c}{4}          & \multicolumn{1}{c}{4+}         & \multicolumn{1}{c}{5}          \\ \hline
GWESP (.1) Class 1         & \textbf{0.131}  & \textbf{0.330}  & 0.024                          & \textbf{0.102}  & \textbf{0.177}  & \textbf{0.672}  & \textit{-0.465} & \textit{-0.512} & 0.037                          \\
GWESP (.1) Class 2         & -                              & -                              & -                              & -                              & -                              & -                              & \textit{1.675}  & \textit{1.930}  & -                              \\
Mutual                     & 0.003                          & 0.008                          & 0.000                          & -0.004                         & 0.003                          & -0.003                         & 0.002                          & 0.001                          & 0.002                          \\
Gender Match               & -0.029                         & -0.077                         & -0.006                         & -0.017                         & -0.040                         & \textbf{-0.154} & -0.073                         & -0.088                         & -0.002                         \\
Ethnicity Match            & -0.065                         &\textbf{ -0.155} & 0.005                          & -0.040                         & -0.087                         & \textbf{-0.263} & \textbf{-0.148} & \textbf{-0.176} & 0.014                          \\
Sender Tobacco Use         & \textbf{0.392}  & \textbf{0.608}  & \textbf{0.208}  & \textbf{0.531}  & \textbf{0.608}  & \textbf{1.085}  & \textbf{0.385}  & \textbf{0.485}  & \textbf{-0.100} \\
Receiver Tobacco Use       & \textbf{-0.118} & \textbf{-0.209} & 0.009                          & -0.091                         & \textbf{-0.182} & \textbf{-0.436} & \textbf{-0.145} & \textbf{-0.200} & \textbf{0.164}  \\
Sender Anti Social         & \textbf{-0.458} & \textbf{-0.783} & \textbf{-0.158} & \textbf{-0.363} & \textbf{-0.572} & \textbf{-1.002} & \textbf{-0.688} & \textbf{-0.823} & -0.013                         \\
Receiver Anti Social       & \textbf{-0.108} & \textbf{-0.175} & -0.042                         & \textbf{-0.129} & -0.088                         & \textbf{-0.367} & \textbf{-0.138} & \textbf{-0.188} & -0.017                         \\
Sender MJ                  & \textbf{-0.291} & \textbf{-0.524} & -0.094                         & -0.085                         & \textbf{-0.415} & \textbf{-0.567} & \textbf{-0.430} &\textbf{-0.530} & -0.009                         \\
Receiver MJ                & -0.033                         & -0.043                         & 0.019                          & -0.010                         & -0.057                         & -0.043                         & -0.005                         & -0.014                         & -0.005                         \\
Receiver Alcohol Use       & \textbf{-0.100} & \textbf{-0.450} & \textbf{-0.250} & \textbf{-1.275} & \textbf{-0.350} & \textbf{-0.650} & \textbf{-0.375} & \textbf{-0.525} & -0.025                         \\
Abs Diff Tobacco Use       & 0.038                          & 0.019                          & -0.010                         & 0.010                          & 0.043                          & 0.024                          & -0.048                         & -0.038                         & -0.076                         \\
Abs Diff MJ                & -0.018                         & -0.064                         & 0.013                          & -0.003                         & -0.031                         & -0.108 & -0.046                         & -0.064                         & -0.005                         \\
Abs Diff Anti Social       & \textbf{-1.525} & \textbf{-2.388} &\textbf{ -0.638} & \textbf{-1.913} & \textbf{-2.113} & \textbf{-3.688} & \textbf{-2.063} & \textbf{-2.413} & -0.025 \\
Abs Diff Alcohol Use       & \textbf{0.100}  & \textbf{-0.200} & \textbf{-0.800} & \textbf{-5.100} & \textbf{-1.300} &\textbf{ -3.400} &\textbf{ 0.100}  & \textbf{-0.100} & \textbf{0.100}  \\
Sender Alcohol Use Class 1 & \textbf{0.817}  & \textbf{1.333}  & -                              & -                              & -                              & -                              & \textbf{0.983}  & \textbf{1.250}  & -0.017                         \\
Sender Alcohol Use Class 2 & -                              & -                              & \textit{-0.416} & \textit{-0.533} & \textit{-0.728} & \textit{-0.836} & -                              & -                              & -                              \\
Edges Class 1              & \textit{0.219}  & \textit{0.421}  & 0.016                          & 0.055                          & \textit{0.244}  & \textit{0.566}  & 0,055                          & 0.053                          & 0.012                          \\
Edges Class 2              & 0.004                          & 0.026                          & -                              & -                              & 0.024                          & 0.071                          & -                              & -                              & -                              \\ \hline
\end{tabular}
}
\begin{tablenotes}
\item MJ: Marijuana, Abs Diff: Absolute Difference
\item \textbf{Bold text} indicates relative bias in homogeneous parameters above .1, \textit{italic text} indicate relative bias in heterogeneous parameters above .1.
\end{tablenotes}
\end{table}

As expected, all of the heterogeneous conditions (1 through 4+) demonstrated large relative bias in many of the homogeneous parameters (Highlighted in Table \ref{homorel} in bold) and increased raw bias (See Table A4). Five effects that had consistent relative bias above .1 in all heterogeneous conditions were Sender Tobacco Use, Sender Anti Social, Receiver Alcohol Use, Absolute Difference in Anti Social and Absolute Difference in Alcohol Use. It should be noted that the Absolute Difference in Anti Social and Absolute Difference in Alcohol Use both had true parameters close to 0 (-.08 and -.1 respectively), therefore a certain amount of relative bias is expected. As shown in Table A4, there is increased raw bias in these parameter estimates as one increases the difference between the latent classes.

Another result of note is the pattern of bias in the GWESP term for Conditions 1 through 3+. With the exception of Condition 2, the relative biases in the estimate of the GWESP term are above .1. This suggests that the GWESP term is sensitive to misspecification of other heterogeneous effects as homogeneous, and it responds to misspecification by increasing in magnitude. Intuitively, this is due to transitivity effects being intrinsically confounded with covariate effects. An edge might form because of a similarity on behavior, or due to the influence of shared partners, and if the effect of similarity on behavior is underestimated, the effect of shared partners might be overestimated to compensate. This bias in a structural effect is not replicated in the mutuality parameter, which suggests that mutuality is robust to misspecifications in the rest of the model.

As expected, there was bias in the parameter estimates that were truly heterogeneous (italicized in Table \ref{homorel}), as it would be impossible to correctly recover the true parameter estimates that were heterogeneous with a single parameter that is homogeneous. Note no relative bias was calculated for the Sender Alcohol Use of Class 1 for Conditions 2 through 3+ as the true value of the parameter was 0. As for Condition 5, while there was some cases of relative bias being higher than .1 (Sender and Receiver Tobacco Use, and Absolute Difference in Alcohol Use) the rest of the parameters were recovered well. This was expected, as for Condition 5 a homogeneous model is the true model.

In summary, Tables \ref{homorel} suggest that fitting homogeneous models to networks that are heterogeneous results in unpredictable patterns of high bias in both parameters that are supposed to be heterogeneous, as well as parameters that are indeed homogeneous.

\subsection{Analysis 2: Recovery of Parameters in Heterogeneous Data using SRFM-ERGMs}

Here we present the relative bias in parameters estimated with correctly specified heterogeneous models, as well as relative bias in parameters estimated with a heterogeneous model applied to homogeneous data. The simulated datasets are the same as used in Analysis 1, with the difference being in the models estimated.

\begin{table}[H]
\centering
\caption{Relative Bias in the Parameter Estimates For Heterogeneous Models, calculated using Mean Raw Bias  over True Parameter Value. Results indicate modest levels of bias in homogeneous parameters. This bias is in line with expected bias from the approximate MLE estimation}
\label{RelB}
\resizebox{\textwidth}{!}{%
\begin{tabular}{lrrrrrrrrr}
\hline
Condition                  & \multicolumn{1}{c}{1}          & \multicolumn{1}{c}{1+}         & \multicolumn{1}{c}{2}          & \multicolumn{1}{c}{2+}         & \multicolumn{1}{c}{3}          & \multicolumn{1}{c}{3+}        & \multicolumn{1}{c}{4}          & \multicolumn{1}{c}{4+}         & \multicolumn{1}{c}{5}          \\ \hline
GWESP (.1) Class 1         & 0.027                          & 0.009                          & 0.013                          & 0.025                          & 0.030                          & 0.010                         & \textit{-0.108} & -0.091                         & 0.048                          \\
GWESP (.1) Class 2         & -                              & -                              & -                              & -                              & -                              & -                             & \textit{0.107}  & \textit{0.228}  & -                              \\
Mutual                     & 0.000                          & 0.000                          & 0.000                          & -0.001                         & -0.001                         & -0.001                        & 0.000                          & 0.000                          & 0.001                          \\
Gender Match               & -0.002                         & 0.006                          & -0.003                         & 0.002                          & -0.003                         & 0.001                         & -0.013                         & -0.016                         & 0.003                          \\
Ethnicity Match            & -0.004                         & 0.010                          & 0.005                          & -0.002                         & -0.005                         & 0.014                         & -0.023                         & -0.025                         & 0.017                          \\
Sender Tobacco Use         & \textbf{0.332}  & \textbf{0.340}  & -0.013                         & -0.088                         & \textbf{0.175}  & 0.032                         & \textbf{0.191}  & \textbf{0.240}  & -0.035                         \\
Receiver Tobacco Use       & 0.032                          & 0.099                          & \textbf{0.131}                          & \textbf{0.198}  & \textbf{0.145}  & \textbf{0.188}                         & 0.063                          & 0.060                          & \textbf{0.167}                          \\
Sender Anti Social         & \textbf{-0.285} & -0.071                         & \textbf{0.108}  & \textbf{0.227}  & \textbf{-0.201} & -0.031                        & \textbf{-0.272} & \textbf{-0.182} & -0.072                         \\
Receiver Anti Social       & -0.059                         & -0.052                         & 0.007                          & 0.049                          & 0.007                          & -0.051                        & -0.034                         & -0.058                         & -0.012                         \\
Sender MJ                  & \textbf{-0.180} & -0.067                         & -0.090                         & 0.017                          & \textbf{-0.196} & -0.071                        & \textbf{-0.230} & \textbf{-0.216} & 0.016                          \\
Receiver MJ                & -0.010                         & 0.010                          & 0.029                          & 0.010                          & -0.043                         & -0.065                        & 0.007                          & -0.003                         & 0.009                          \\
Receiver Alcohol Use       & 0.083                          & -0.011                         & \textbf{-0.145}                         &\textbf{-0.263}                         & -0.012                         & \textbf{0.198} & -0.036                         &\textbf{-0.100}                         & -0.013                         \\
Abs Diff Tobacco Use       & -0.013                         & -0.039                         & -0.066                         & -0.100                         & -0.073                         & -0.093                        & -0.051                         & -0.054                         & -0.074                         \\
Abs Diff MJ                & 0.004                          & 0.002                          & 0.019                          & -0.001                         & -0.014                         & -0.026                        & 0.002                          & -0.009                         & 0.003                          \\
Abs Diff Anti Social       & \textbf{-0.425} & \textbf{-0.165} & 0.085                          & \textbf{0.517}  & \textbf{-0.353} & 0.011                         & \textbf{-0.495} & \textbf{-0.486} & -0.032                         \\
Abs Diff Alcohol Use       & \textbf{0.341}  & \textbf{0.137}  & \textbf{-0.570} & \textbf{-0.950} & -0.055                         & \textbf{0.638} & \textbf{0.232}  & 0.036                          & \textbf{0.141}                          \\
Sender Alcohol Use Class 1 & \textbf{0.291}  & 0.075                          & NA                             & NA                             & NA                             & NA                            & \textbf{0.502}  & \textbf{0.388}  & \textbf{-0.110}                         \\
Sender Alcohol Use Class 2 & -                              & -                              & 0.050                          & -0.013                         &  \textit{-0.163}                         & -0.003                        & -                              & -                              & \textbf{-0.237} \\
Edges Class 1              & 0.055                          & 0.015                          & -0.001                         & -0.004                         & 0.049                          & 0.008                         & 0.020                          & 0.009                          & -0.020                         \\
Edges Class 2              & 0.031                          & 0.006                          & -                              & -                              & 0.028                          & -0.049                        & -                              & -                              & 0.064                          \\ \hline
\end{tabular}
}
\begin{tablenotes}
\item MJ: Marijuana, Abs Diff: Absolute Difference, NAs due to true effect being 0.
\item \textbf{Bold text} indicates relative bias in homogeneous parameters above .1, \textbf{italic text} indicates relative bias in heterogeneous parameters above .1.
\end{tablenotes}
\end{table}

In terms of bias in the parameter estimates, recovery of heterogeneous parameters was far superior than seen in the results for the homogeneous models discussed above. Importantly, the homogeneous effects (such as gender match, ethnicity match, etc.) also evidenced less bias (see Table A-2). Indeed, very few parameters had greater than .1 absolute raw bias or greater than .1 relative bias. 

While modeling heterogeneity greatly improved the degree of bias, there were a few parameters that still evidenced suboptimal parameter estimation. The greatest bias was seen in Conditions 4 and 4+, where the relative bias for the recovery of the lower GWESP term (GWESP Class 2) had a true value of .2 in both conditions. The relative bias here was .107 for Condition 4 and .228 for Condition 4+. However, the raw bias (contained in Supplementary Material Table A2) in terms of magnitude was quite small, .021 and .046 respectively. This level of bias is not relevant to researchers analyzing empirical data, and is likely due to the GWESP term being difficult to estimate in general. Of note, the GWESP estimates contained very little relative or absolute bias in the other conditions once heterogeneity was modeled. 

The one other heterogeneous term that had a relative bias of above .1 was that of Sender Alcohol Use in Class 2 for Condition 3. The relative bias for this parameter was -.163, with a raw bias of .041. The true parameter value was -.25, which suggests that while the relative bias was high, the actual level of raw bias was comparatively low. Additionally, compared to the homogeneous results (where 4 conditions evidenced high bias on this effect), that only one condition in the heterogeneous modeling framework evidenced bias is a substantial improvement.  

A few homogeneous effects contained bias. The Sender Anti Social effect for conditions 1, 2+, 3, 4, and 4+ all have relative bias magnitude greater than .1. This effect was held homogeneous across all conditions and had the highest magnitude of the homogeneous effects (-.64). The patterning of bias across all conditions was not consistent, and this likely indicates that the effect was being disrupted by the effects of the structural parameters such as GWESP. 

Additionally, with Sender Tobacco Use, several conditions had relative bias greater than .1. However, the true effect of Sender Tobacco Use was small (.13) in all conditions, which can lead to small levels of raw bias translating into large relative bias. This occurrence of high relative biases for very small true parameters is expected and most notably occurs for the Absolute Difference of Anti-Social Behavior, Sender Alcohol use (for models with homogeneous Sender Alcohol Use), and Absolute Difference in Alcohol Use. Furthermore, for Conditions 2, 2+, 3 and 3+ the true effect of Sender Alcohol Use in Class 1 was 0, which would lead to undefined relative bias for any level of raw bias. 

As for Condition 5, recovery of homogeneous parameters was quite good, with no raw bias being greater than .1, and very few relative biases greater than .1. This suggests that the misspecification of the latent class structure did not lead to systematic bias in the homogeneous parameters. Additionally, the spread of the latent class parameters around the true homogeneous parameters was quite reasonable, with Sender Alcohol Use having a true effect of .06, and the latent classes returning an effect of .053 and .046, while the Edge effect is 3.85 while the latent classes returned effects of 3.773 and 4.097 respectively.

In summary, Table \ref{RelB} suggest that the sender latent class models fit to heterogeneous data do a reasonable job of recovering the model parameters. Importantly, the results are far less biased than those seen in the results when homogeneous models are fit to the heterogeneous data sets. Bias in the parameter estimates is in part due to ERG modeling being approximate even when using the MCMC-MLE. A degree of bias is to be expected. That being said, the fact that bias still remained even when fitting the generating model to data suggests that researchers should have stringent criteria for interpreting results from latent class ERGMs, interpreting both effect size as well as significance level.

As for estimated standard errors of the estimates and the standard deviations across the simulation trials for every homogeneous parameter, the estimated standard error of the homogeneous parameter and the standard deviation of the parameter across iterations within a condition were similar. This suggests that the estimates for the standard errors of homogeneous parameters are accurate. However, for the heterogeneous parameters the estimated standard error is considerably smaller than the simulation standard deviation (Table A6 in supplementary material). This is in part due to error in classification across the simulation trials. If there are simulation trials where the individuals are misclassified then the estimates of the heterogeneous effects will be more variable than if in every trial the latent class labels were recovered with perfect fidelity. Additionally, the estimated standard errors assume that the latent class assignment is the true assignment, and are not adjusted for uncertainty in class assignment. This is a reflection of the classification likelihood EM algorithm in use in the estimation. 

\subsubsection{Analysis 2: Latent Class Label Recovery with SRFM-ERGMs}

Table \ref{Rands} shows the mean Adjusted Rand Indices for latent class recovery for every individual, as well as for only individuals who use alcohol.  

\begin{table}[H]
\centering
\caption{Mean Adjusted Rand Indices (Standard Deviations computed from the sample) computed across 500 trials within each condition. Results indicate best recovery occurs in the + conditions, with higher class separation. Structural effects such as GWESP are also recovered well.
}
\label{Rands}
\begin{tabular}{ccc}
\hline
\multicolumn{1}{c}{Condition}  & \multicolumn{1}{c}{ARI}  & \multicolumn{1}{c}{ARI Alcohol Users} \\ \hline
1                                      & 0.541 (.126)                                     & 0.532 (.179)                          \\
1+                                      & 0.924 (.051)                                   & 0.89 (.104)                           \\
2                                        & 0.06 (.120)                                   & 0.272 (.137)                          \\
2+                                      & 0.1 (.178)                                     & 0.703 (.108)                          \\
3                                      & 0.629 (.120)                                     & 0.798 (.137)                          \\
3+                                   & 0.959 (.034)                                  & 0.997 (.014)                          \\
4                                     & 0.667 (.096)                                   & 0.678 (.153)                          \\
4+                                    & 0.862 (.069)                                 & 0.848 (.113)                          \\ \hline
\end{tabular}

\end{table}
Adjusted Rand Indices (ARI) were computed as the empirical mean and empirical standard deviation over every trial in a given condition. For class recovery, the mean ARI contained in Table \ref{Rands} indicate that across all simulation conditions, increasing the difference in parameter weights between the latent classes leads to increased recovery of the latent classes. However, for conditions with smaller class separation (Conditions 1, 2, 3, 4), the recovery of the latent classes was not optimal.  For most conditions, the greatest improvement in recovery of the latent class labels occurred only when looking at alcohol users.  The exception being conditions 2 and 2+ where even larger differences between the latent classes, recovery of the class labels for alcohol users is still not as good as for other conditions (ARI of .703 for Condition 2+). This is due to the latent class being defined solely on individuals who used alcohol, a small subset of the sample, as opposed to other conditions that had the latent classes also be defined by structural parameters such as edges or GWESP. Similarly, as the latent classes were defined only on sender alcohol use for Conditions 2 and 2+, individuals who did not use alcohol had no information to inform their theoretical class membership. This in turn explains the very low ARI values for Conditions 2 (.06) and 2+ (.1) across the whole sample, and the improvement in ARI values when looking at only individuals who used alcohol. These result emphasize the need for researchers to take care when specifying which effects the latent classes vary in. If one were to specify heterogeneous parameters for a covariate that only a small subset of the network has a non-zero value on, then the latent class solution will only be valid for that subset. 

One set of results of note is that Condition 4 and 4+ had remarkably good recovery of the latent class labels, even at the lower class difference (ARI of .667 and .862  for Condition 4 and 4+ respectively). This suggests that heterogeneities in the GWESP term are reflected strongly in observed network structure. Finally, standard deviations of the ARIs suggest that there was variability in recovery rates due to sampling fluctuations. This variability can lead, particularly in conditions with low class separation, to very poor rates of classification. As one would expect, there is more variability in the ARIs the lower the average ARI becomes.

\subsection{Analysis 3: Effect of Sample Size on Latent Class Recovery}

\begin{figure}[H]

\centering
\includegraphics[scale =.8]{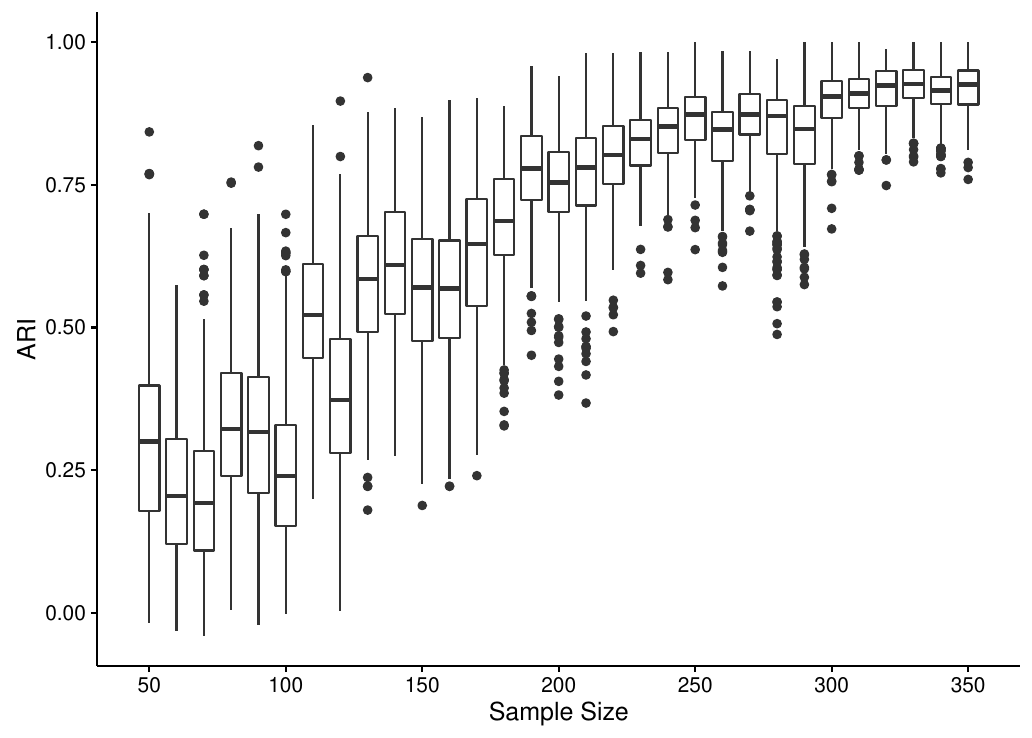} 
\caption{ARI by Sample Size for Condition 3 simulations}
\label{Cond3}
\end{figure}

\begin{figure}[H]

\centering
\includegraphics[scale =.8]{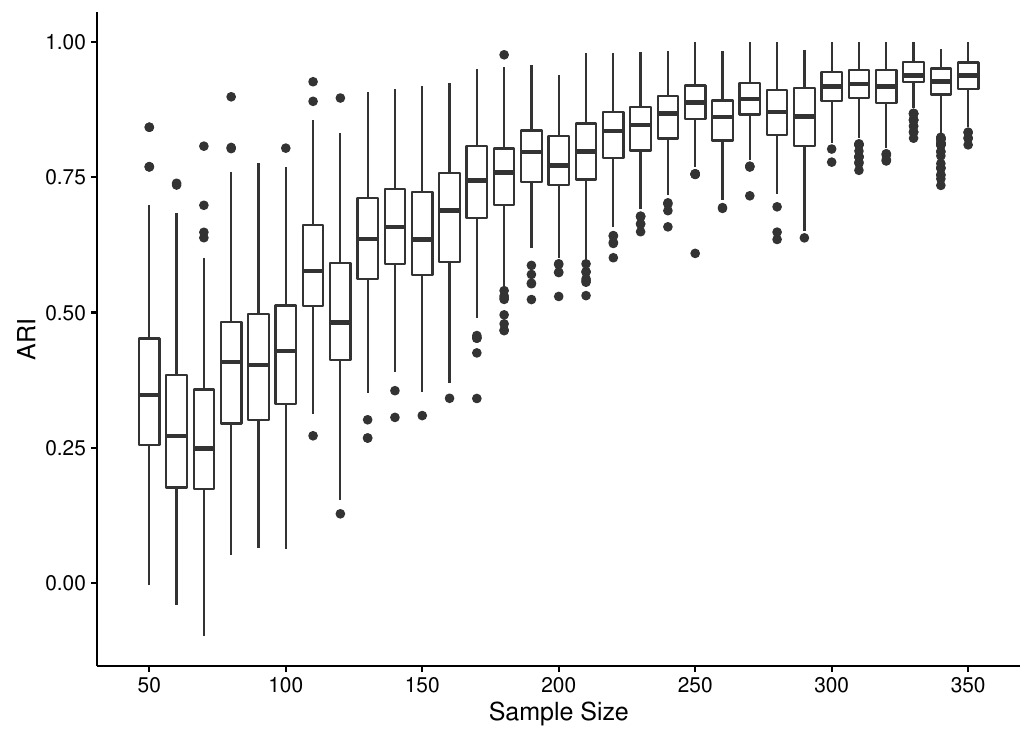} 
\caption{ARI by Sample Size for Condition 4 simulations}
\label{Cond4}
\end{figure}

Figure \ref{Cond3} and \ref{Cond4} shows the distribution of Adjusted Rand Indices as a function of increased sample size. Classification accuracy improves as one would expect as a function of sample size. The lack of a smooth curve in classification accuracy is a result of the simulation procedures. While 500 networks were simulated per sample size target, only one covariate set was generated for a given sample size. This allows the variation in classification to be a function mainly of the network, rather than the covariate set, however results in some counter intuitive results in the classification, such as the jumps in classification accuracy exhibited for several sample sizes. The overall shape of the curve suggests that the estimates of the latent class labels are improving with sample size. 

\section{Discussion}

This manuscript presented the Sender/Receiver Finite Mixture ERGM, provided a estimation routine, presented simulations that demonstrated its ability to detect latent class structures in networks, and demonstrated the consequences of failing to model latent class structures in network. Additionally, this manuscript presented an empirical example rooted in adolescent peer networks that demonstrated this models ability to obtain reasonable results from real, rather than simulated data. 

There are several findings that can be gleaned from the simulations. First, recovery of the latent class structures is sensitive to the difference between the latent classes, as demonstrated by the increased difference conditions in the simulations. This is to be expected; however, caution is advised when using these models on empirical data as the models will return solutions regardless of underlying differences between the latent classes. One solution to this problem is examining the {\it a posteriori} probabilities of class membership given the returned set of parameter estimates. If there is a large amount of uncertainty in the classifications of individuals, then the model should be reassessed. Simulation results suggest that the classification probabilities should be quite high for correctly classified individuals. Additionally, penalized likelihood criteria, such as AIC or BIC, or entropy based criteria (e.g. NEC; \cite{Biernacki1999,Celeux1996}, Conditional Classification Criteria; \cite{Baudry2015}) can provide guidance for determining the presence and the number of latent classes in empirical data. In particular, entropy based criteria are particularly well suited for the classification approach used here, as those criteria will be sensitive to class separation, which, as noted previously, the classification EM attempts to maximize. Additionally, another advantage of the hard classification approach taken here is that poorly separated classes will tend to collapse, making this estimation approach appropriate when there is \textit{a priori} reasons to assume the existence of distinct classes.

One key point that can be taken away from this study is that caution is warranted when analyzing data with homogeneous models. As shown in the simulations, mis-specifying structural parameters such as edges or GWESP can result in wide-ranging bias of many other model parameters, where this bias is generally unpredictable in its direction. Additional caution is warranted when looking at the effects of covariates as a mis-specification of one covariate effect (such as a sender effect) as homogeneous when really there are heterogeneous effects leads to bias in other effects of the same covariate (such as absolute difference). This is particularly concerning as the direction of the bias is unpredictable, and therefore mis-specification could lead to spurious findings. 

As another finding of particular interest, mis-specifying a heterogeneous model to a truly homogeneous network did not result in any undue bias, other than in the mis-specified terms. Even with the mis-specified terms the differences in effect size were small. This is in part due to the nature of the classification likelihood, in that it tends to collapse components that are ill separated. This suggests that researchers can use heterogeneous models as an exploratory method, without worry that a heterogeneous mis-specification of a homogeneous network would result in biased estimates. 

The empirical example offers support for this method in answering substantive questions. Here, that question was: does the effect of alcohol use on the number of friends one nominates differ between individuals? With this sample, there appears to be at least some level of heterogeneity in the effect of alcohol use on the number of friends an individual nominates. Furthermore, by modeling heterogeneity, several effects were clarified that would have not been significant if a homogeneous model was run. These findings further emphasize the importance of modeling individual differences in network data.

There are several limitations to the SRFM-ERGMs. First, the estimation technique for SRFM-ERGMs described here requires a directed network.  Secondly, the estimation method require the analyst to make a choice to model either Sender latent classes, or Receiver latent classes, but not both. This restricts the type of heterogeneity that an analyst can model. Finally, as developed the Sender/Receiver Latent Class ERGM is only for cross-sectional data.

These limitations open up several new directions for future work. The next step would be extending these models to fit general finite mixture ERGMs, rather than restricting the latent class definition to be sender or receiver. A general finite mixture ERGM would resemble the ERGM models proposed by Schweinberger and Handcock \citeyear{schweinbergerlocal2015}, however would not focus on latent community detection (though they would be capable of it). Additionally, a general latent class ERGM would be able to model undirected networks. Another extension would be to develop a longitudinal version. Adapting the separable temporal ERGMs \citep{Krivitsky2014} which can model peer selection processes over time, would allow analysts to investigate heterogeneous peer selection processes. However, it should be noted that the computational difficulties induced by allowing both Sender and Receiver effects will not be alleviated by any estimation technique, and this will limit more general models applicability to very large networks. Finally, more work needs to be done on how to determine the number of classes when using a classification EM approach.  

Overall, this manuscript introduces a new way of modeling network data and a different definition of heterogeneity on networks that could better serve substantive researchers studying a variety of heterogeneous and latent network-rooted behaviors and traits, such as in adolescent peer influence of substance abuse (eg., Brechwald \& Prinstein, 2011), in clinical domains such as autism research (eg., \citep{gilmanrare2011}), and also in social, cognitive, and affective neuroscience \citep{wagera2015}. Further work is needed to refine and expand the latent class ERGM so that psychological and social scientists can advance our understanding of ideographic variation in network settings.

\vspace{\fill}\pagebreak

\bibliographystyle{apalike}
\bibliography{Biblio}




%




\end{document}